\documentclass[aps,jmp,amsmath,amssymb,reprint]{revtex4-1}

\usepackage{graphicx}
\usepackage{color}

\begin{document}

\title{Recovery of hexagonal Si-IV nanowires from extreme GPa pressure}

\author{Bennett E. Smith}
\affiliation{Department of Chemistry, University of Washington}

\author{Xuezhe Zhou}
\author{Paden B. Roder}
\affiliation{Department of Materials Science \& Engineering, University of Washington}

\author{Evan H. Abramson}
\affiliation{Earth \& Space Sciences Department, University of Washington}

\author{Peter J. Pauzauskie}
\email[]{peterpz@uw.edu}
\affiliation{Department of Materials Science \& Engineering, University of Washington, \& Fundamental \& Computational Sciences Directorate, Pacific Northwest National Laboratory}

\begin{abstract}
We use Raman spectroscopy in tandem with transmission electron microscopy and DFT simulations to show that extreme (GPa) pressure converts the phase of silicon nanowires from cubic (Si-I) to hexagonal (Si-IV) while preserving the nanowire's cylindrical morphology.  In situ Raman scattering of the TO mode demonstrates the high-pressure Si-I to Si-II phase transition near 9 GPa.  Raman signal of the TO phonon shows a decrease in intensity in the range 9 to 14 GPa. Then, at 17 GPa, it is no longer detectable, indicating a second phase change (Si-II to Si-V) in the 14 to 17 GPa range.  Recovery of exotic phases in individual silicon nanowires from diamond anvil cell experiments reaching 17 GPa is also shown.  Raman measurements indicate Si-IV as the dominant phase in pressurized nanowires after decompression.  Transmission electron microscopy and electron diffraction confirm crystalline Si-IV domains in individual nanowires.  Computational electromagnetic simulations suggest that heating from the Raman laser probe is negligible and that near-hydrostatic pressure is the primary driving force for the formation of hexagonal silicon nanowires.
\end{abstract}

\maketitle

Silicon is the second most abundant element in the Earth's crust \cite{lut12} and the foundation of the modern electronics industry.   It is used for integrated circuits in information technology and as an energy conversion material in photovoltaics.  Unfortunately, one of the biggest drawbacks for silicon's use in solar energy conversion is its indirect band gap.  Theoretical \cite{rod15, wip13} and experimental \cite{fon07, zha99, kim15} efforts are looking at the properties of exotic phases of silicon and their potential as improved photovoltaic (PV) absorbers.

The phase diagram of silicon \cite{vor03} reveals several polytypes at elevated pressures.  At a pressure of $\sim$11 GPa, Si-I begins to transition to Si-II which has a body-centered tetragonal crystal structure and metallic electronic structure \cite{dyu78}.  As pressure increases past approximately 15 GPa, Si-V begins to emerge with a primitive hexagonal phase \cite{hu84, hu86, tol95}.  But neither Si-II nor Si-V are stable at atmospheric pressure and, therefore, have not been observed experimentally outside of high pressures.  Si-III (body-centered cubic) and Si-IV (diamond hexagonal), however, are stable at atmospheric pressure and have been synthetisized \cite{fon07, fab13, fab14} as well as recovered from high-pressure phase transitions \cite{bes87, wei89}.  While Si-III is a semimetal \cite{bes87} and could have applications in electronics, Si-IV is a semiconductor with a reported indirect band gap near 0.8-0.9 eV and direct transition at 1.5 eV \cite{fab14}.  The direct transition for Si-IV makes it appealing for PV applications due to higher absorption efficiency in the visible spectrum.
\begin{figure}
\includegraphics[width=3.3in,bb=0 0 9.1in 3.08in]{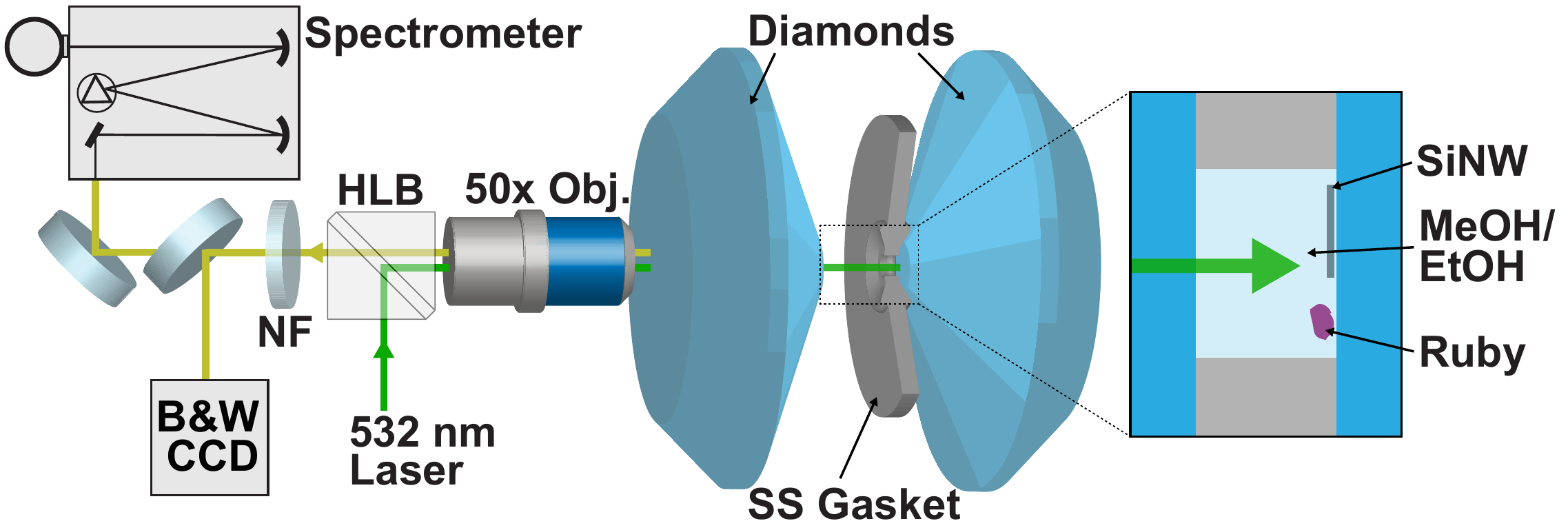}
\caption{Schematic of the diamond anvil cell (DAC) and components for Raman scattering measurements.  A holographic laser bandpass (HLB) filter is used to pass the 532 nm Raman probe into a 50x objective which focused the beam into DAC.  Nanowire Raman scattering and ruby photoluminescence were collected with the same objective and sent to a spectrometer or CCD for imaging.  A 532 nm notch filter (NF) was used to eliminate strong Rayleigh scattering. \label{scheme}}
\end{figure}
Nano-structured silicon can also prove advantageous for light-absorbing materials.  For example, periodic dielectric structures, including photonic crystals (PC), are used to control and confine the movement of photons in two or three dimensions\cite{yab93, joa97}. PCs of silicon can be synthesized using finely controlled methods such as lithography \cite{bir01} and glancing angle deposition \cite{ken03}.  Recently, a 2-D silicon PC was shown to increase absorption efficiency of photovoltaic cells by 31\% when compared with a c-Si film with a distributed Bragg reflector \cite{ber07}. Additionally, analytical theory has shown that morphology-dependent resonances can enhance internal fields in silicon nanowires \cite{rod12}.  Experimental results have demonstrated increased absorption in nanowire-patterned silicon as compared with planar silicon across the visible and near-infrared regions of the electromagnetic spectrum \cite{gar10}.  Although Si-IV nanowires with direct-gap transitions have been synthesized through chemical vapor deposition methods\cite{fab14}, we show here that Si-IV can be recovered in silicon nanowires previously prepared through chemical etching after near-hydrostatic compression up to 17 GPa in a diamond anvil cell (DAC).  These results demonstrate the feasability of designing PCs with cubic silicon and recovering exotic phases after pressurization while maintaining complex morphologies created through lithographic processing.

\begin{figure}[t]
 \includegraphics[width=3in,bb=0 0 4.15in 5.47in]{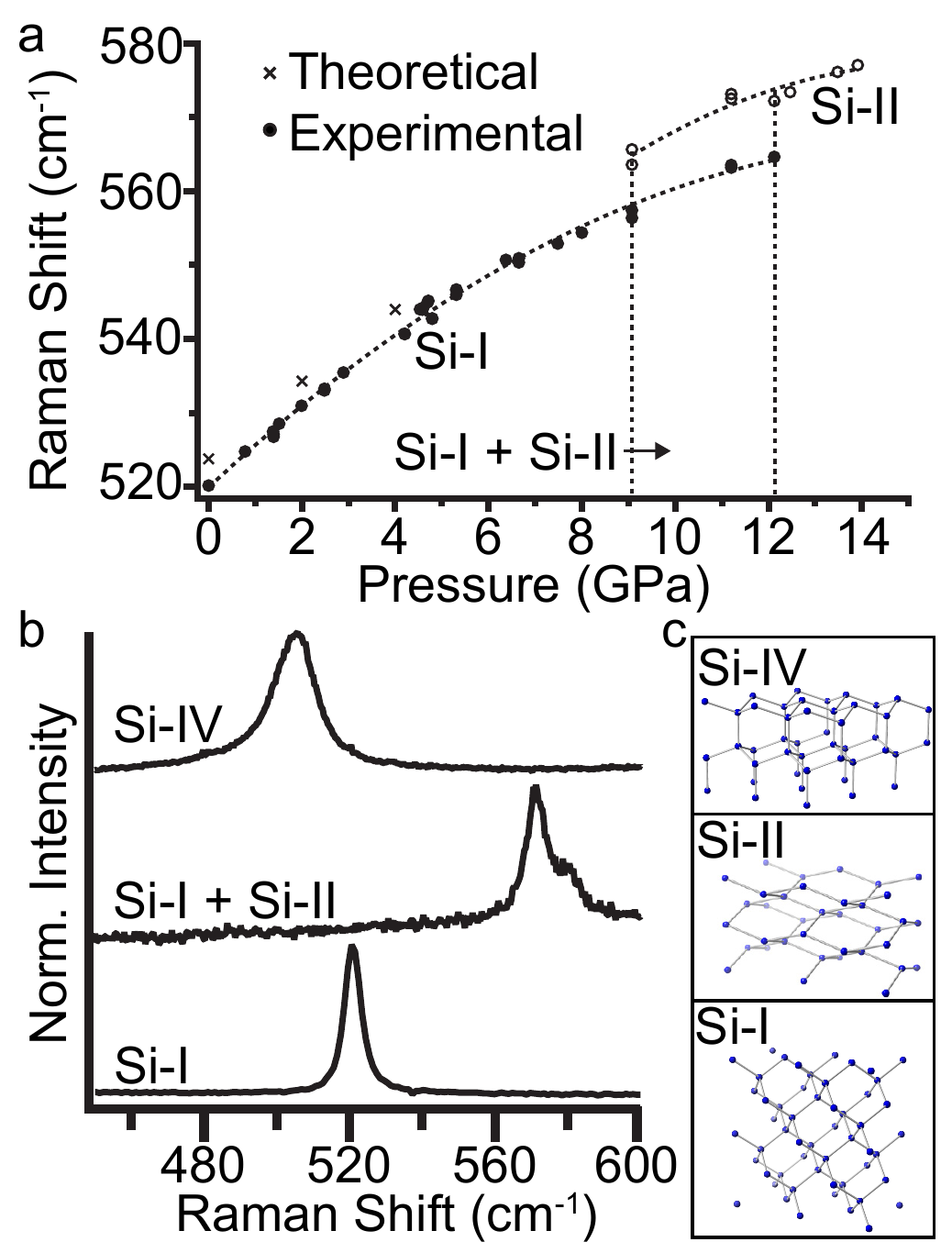}
 \caption{(a) In-situ Raman scattering from 12 individual SiNWs under compression in the DAC shows the onset of the Si-I to Si-II phase transition near 9 GPa with a complete transition to Si-II at 12.3 GPa. DFT simulations of the Si-I LTO mode agree well with experimental results in the range of 0 to 4 GPa.  (b) Raman scattering of the same recovered SiNW from Figure \ref{tem} confirms a Si-IV phase while measurements from a SiNW at a pressure of 12.3 GPa show the existence of two distinct modes.  Raman scattering from an uncompressed SiNW is plotted for comparison.  (c) Crystal structures for the relevant phases of silicon: diamond cubic (Si-I), body-centered tetragonal (Si-II), and diamond hexagonal (Si-IV). \label{raman}}
\end{figure}
Silicon nanowires (SiNWs) were prepared through metal-assisted chemical etching (MACE) \cite{hua11}. A silicon wafer with $<$111$>$ orientation, doped with boron to a resistivity of 11 $\Omega$ cm, was immersed in a solution of 1:1 volume ratio of 10 M HF:0.04 M AgNO$_3$ for 3 hours.  The etched wafer was then immersed in a 1:1 volume ratio solution of 30\% (v/v) NH$_4$OH:28\% (v/v) H$_2$O$_2$ which has been shown to remove any residual silver particles \cite{smi15}.  The resulting nanowire array was sonicated to suspend the nanowires in deionized H$_2$O.  Pressure modulation was achieved using a Boehler-Almax plate DAC and diamonds with 0.3 mm culets.  A tungsten probe was used to transfer nanowires dried from the suspension onto the diamond anvil culet along with micrometer-scale ruby grains which were used to monitor the pressure inside the cell \cite{mao86}.  A rhenium gasket was used as a spacer after it was dimpled to a thickness of 50 $\mu$m and a hole with 150 $\mu$m diameter was drilled with an electrostatic-discharge machine (Hylozoic Products, Seattle). The cell was then sealed and pressurized initially using either a 4:1 volume ratio or methanol:ethanol mixture or cryogenically loaded argon as a near-hydrostatic pressure transfer medium.  By gradually tightening the cell, the diamond culets were advanced closer to each other, thereby increasing pressure in the cell.

Raman scattering from individual SiNWs at high pressures was observed by focusing a 532 nm laser through a 50x objective into the DAC, as illustrated in Figure \ref{scheme}, to a spot size of $\sim$3 $\mu$m and dispersing the back-scattered signal onto a liquid nitrogen-cooled charge-coupled device (CCD) through a 0.5 m spectrograph with a 2400 l/mm holographic grating and slit-width of 20 $\mu$m giving a resolution of 0.3 cm$^{-1}$.  Raman shift values obtained from spectra of SiNWs were calibrated to $\pm$1 cm$^{-1}$ using Raman scattering spectra from cyclohexane.  

In situ Raman spectra from 12 individual SiNWs at increasing pressures (Fig.\ \ref{raman}a) show a quadratic dependence of the Si-I first-order longitudinal-transverse optical (LTO) phonon with a fit described by the equation
\begin{multline}
\omega_I=520.0\pm0.44\text{ cm$^{-1}$}\\
+(5.88\pm0.17\text{ cm$^{-1}$ GPa$^{-1}$})P\\
+(-0.184\pm0.014\text{ cm$^{-1}$ GPa$^{-2}$})P^2.
\end{multline}
The vibrational frequencies of atoms in a solid are dependent on the volume and can be described by the mode Gr\"uneisen parameter
\begin{equation}
\gamma_i=-\frac{\partial\text{ln}\omega_i}{\partial\text{ln}V}=\frac{1}{\chi_T\omega_i}\frac{\partial \omega_i}{\partial P}
\end{equation} 
where $\omega_i$ is the frequency of the $i$th mode, $V$ is the crystal volume, and $\chi_T$ is the isothermal compressibility.  Using a bulk value \cite{kit05} for $\chi_T=0.01012$ GPa$^{-1}$, the mode Gr\"uneisen parameter is calculated to be 1.12, which is in agreement with other findings \cite{kha13, wei75}.

 \begin{figure}
\includegraphics[width=3in,bb=0 0 290 292]{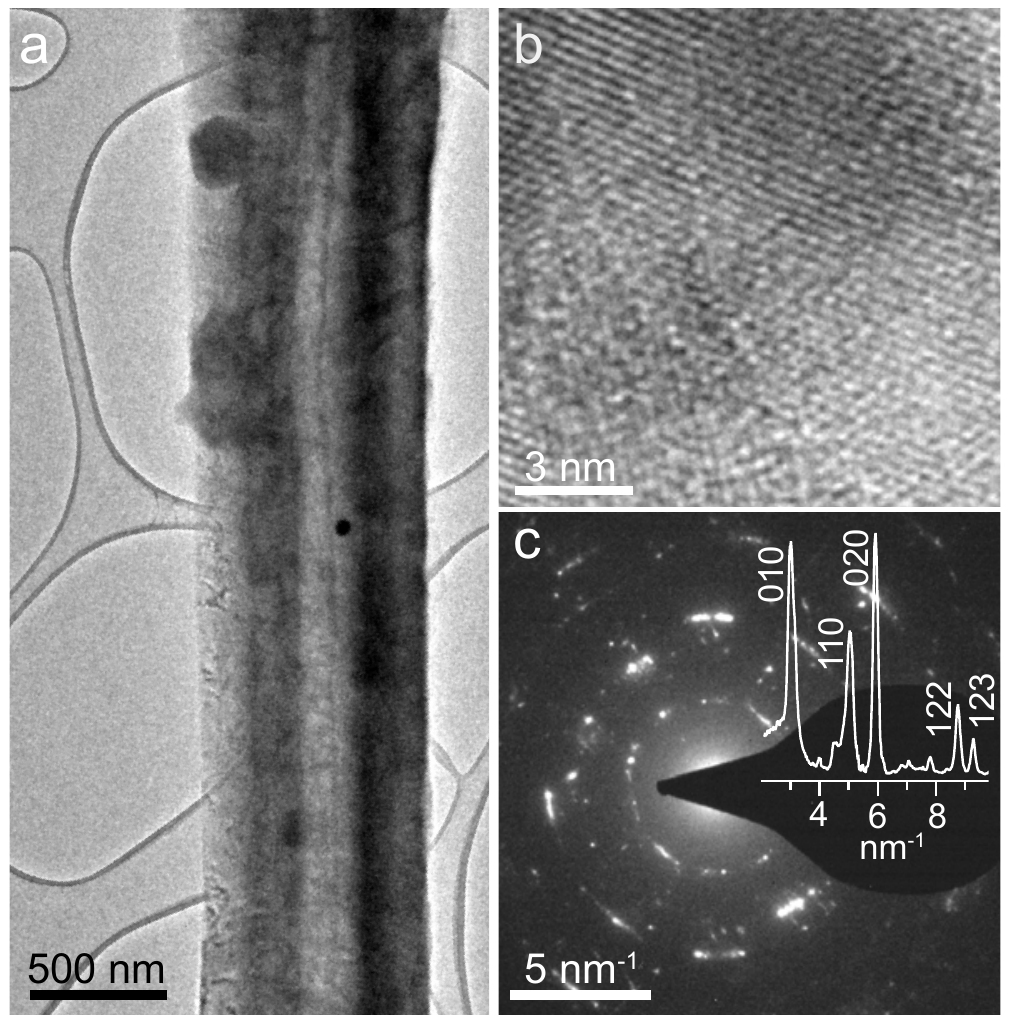}
 \caption{(a) Bright-field transmission electron micrograph of a SiNW recovered from DAC experiments which ultimately reached a pressure of 17 GPa. (b) High resolution TEM of the SiNW shows that a crystalline structure was maintained upon decompression. (c) SAED of the SiNW demonstrates diffraction from multiple domains and the integrated ring pattern can be indexed to a Si-IV phase. \label{tem}}
\end{figure}
Unique to these measurements is the existence of two modes simultaneously (Fig.\ \ref{raman}b) in the pressure range of 9--12 GPa.  Previous reports \cite{dyu78, hu86, vor03} have shown through X-ray measurements that both Si-I (diamond) and Si-II (tetragonal) (Fig.\ \ref{raman}c) are stable at these pressures, however, Raman scattering from Si-II has, to the authors' knowledge, not yet been observed.  The two distinct modes have a separation of approximately 10 cm$^{-1}$ which persists until 12.3 GPa when the signal from the Si-I mode vanishes, suggesting the completion of the Si-I to Si-II phase transition.  The continuity of the new mode past the I-II phase transition and up to 14 GPa indicates that the origin of the mode is indeed from the Si-II phase since it has been shown \cite{hu84, hu86} to be the stable phase across the range of 12--14 GPa.  The next increase in pressure resulted in a sudden jump to 17 GPa, at which point the Raman signal was no longer discernible.  This loss of signal is likely due to another phase transition, Si-II to Si-V (primitive hexagonal), which is expected near 15 GPa \cite{dyu78}.  
Upon decreasing the pressure from the maximum pressure achieved (17 GPa) to 10 GPa it was noted that Raman scattering could not be recovered from any of the SiNWs.  Attempts were made to decrease pressure gradually below 10 GPa to probe for additional phase changes, but further loosening of the DAC tension screws resulted in an uncontrolled pressure release to 1 bar.  However, once atmospheric pressure was restored and the NWs could be probed directly, Raman signal from exotic phases was observed with the dominant signal resulting from the Si-IV phase (Fig.\ \ref{raman}b) which has a diamond hexagonal structure.  Recovery of Si-IV in MACE-prepared SiNWs demonstrates the possibility of also recovering Si-IV in exotic morphologies of Si produced through lithography \cite{bir01} or other methods.

We were also interested in evaluating the predictive capabilities of density functional theory (DFT) for pressure-dependence of Raman shifts and high pressure phase transitions.  Simulations were performed with the Quantum Espresso software package \cite{gia09} using a diamond cubic unit cell with 8 Si atoms and a lattice parameter of 5.43 \AA.  A variable cell relaxation and subsequent phonon calculation were performed across a range of pressures, yielding theoretical Raman shifts for the LTO mode of Si-I.  For the range of 0-4 GPa, the numerically calculated value for $\Delta\omega/\Delta P$ was found to be 5.05 cm$^{-1}$/GPa, which has a difference of less than 2\% from the experimental result.  However, as the simulated pressure was increased past 4 GPa, the disagreement between theory and experiment became significant and no phase change occurred in the simulation even up to pressures of 80 GPa.  The results of these simulations indicate the inaccuracy of DFT for high pressure phase transitions of Si. 

After decompression, recovered SiNWs were then transfered to a lacey carbon transmission electron microscopy (TEM) grid for further structural characterization.  Bright-field TEM images reveal multiple domains within a single nanowire (Fig.\ \ref{tem}a).  High-resolution TEM (Fig.\ \ref{tem}b) shows the crystallinity of one domain as an example while ring patterns from select area electron diffraction (SAED) (Fig.\ \ref{tem}c) over the entire nanowire confirm the existence of multiple phases within a single SiNW where each of the strongest peaks can be indexed to planes from the Si-IV wurtzite phase with unit cell parameters of $a=3.8$ \AA\ and $c=6.27$ \AA.  Secondary peaks suggest domains of either Si-I or Si-III, but the intensities are too weak for conclusive assignment.

\begin{figure}[b]
 \includegraphics[width=3.3in,bb=0 0 10.42in 3.94in]{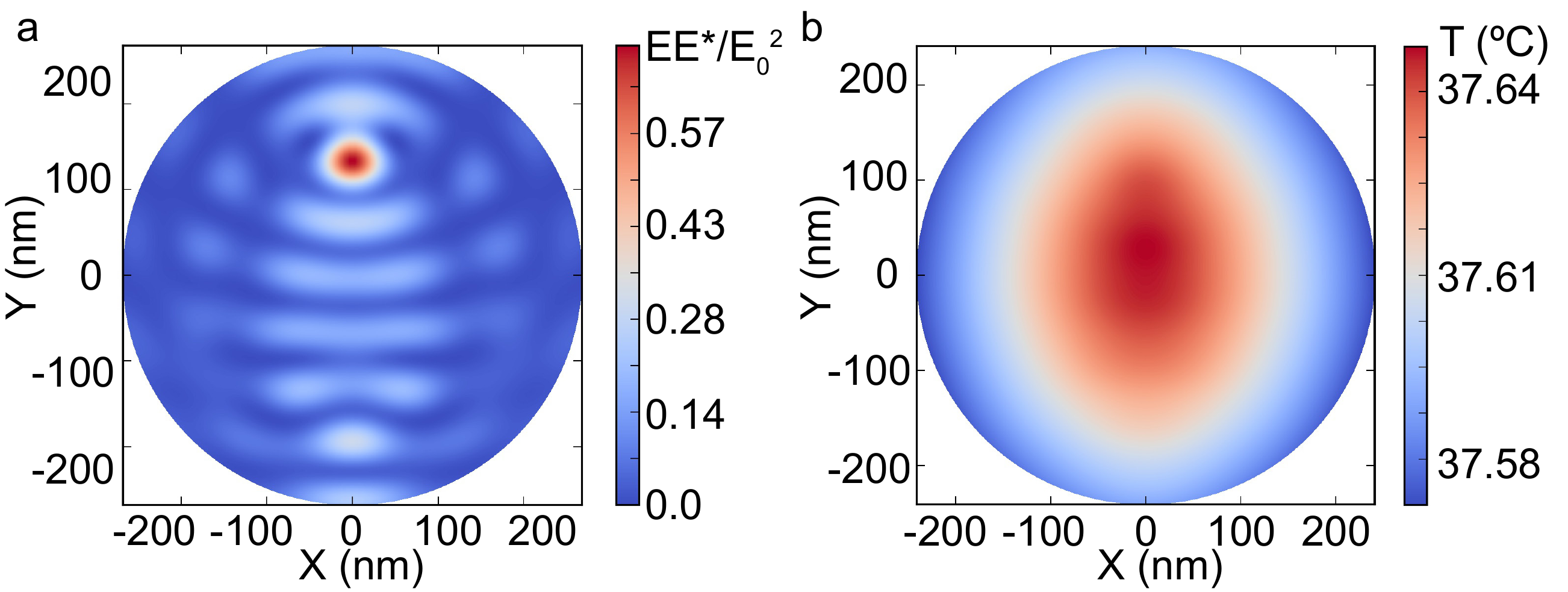}
 \caption{(a) Theoretical, normalized internal electric field for a SiNW with a 483 nm diameter illuminated with a 532 nm laser in solid argon.  (b)  Theoretical temperature profile for the SiNW under 25 kW/cm$^2$. \label{sim}}
\end{figure}
Laser heating was reported by Khachadorian et al.\ \cite{kha11} to influence Raman measurements of SiNWs.  Our results, however, did not show any change in Raman shift or linewidth of the LTO Si mode for laser powers in the range of 1 to 40 mW; although, all Raman measurements conducted at higher pressures near phase transitions ($>$ 8 GPa) were done with a laser power of 5 mW or less.  Furthermore, a custom Python code implementation of analytical theory for laser heating of infinite cylinders \cite{rod12} was used to predict the temperature of a SiNW in a matrix of 4:1 methanol:ethanol with a thermal conductivity of 2 W m$^{-1}$ K$^{-1}$ \cite{hsi15}; this provides an upper bound on the temperature when considering the NW is positioned in contact with one diamond anvil surface.  As shown in Figure \ref{sim}a, the internal field has a maximum that is approximately 70\% of the incident field's magnitude which corresponds to a temperature rise that is less than 13$^\circ$C.  Calculations for cylinders in a solid matrix of argon where the thermal conductivity is 5x higher \cite{gon12} than that of the ethanol:methanol mixture predict a temperature rise of only 2$^\circ$C.  Therefore, pressure is likely the primary mechanism behind the observed phase transition to Si-IV.

In conclusion, silicon is one of the most popular materials in PV devices even though its indirect bandgap limits its conversion efficiency.  Nanowires can be used to enhance internal fields and optical absorption but conversion of Si-I to Si-IV with a direct transition could also significantly increase efficiency in PV devices.  Raman scattering from individual SiNWs up to a pressure of 17 GPa indicates the onset of a pressure-induced phase change from Si-I to Si-II near 9 GPa with complete transtition at 12.3 GPa and Si-II to Si-V transition between 14 and 17 GPa.  We have also recovered NWs at atmospheric pressure which demonstrate polymorphic Si with Si-IV as the dominant phase as evidenced by Raman scattering.  Contrast seen in bright-field TEM images indicate multiple domains while high resolution TEM and SAED confirm that these domains are poly-crystalline and exhibit a Si-IV phase.  It would be of interest to perform compression recovery experiments on 1-D PCs of silicon to learn how the phase will influence the photonic properties of nanostructured Si materials (e.g. photonic crystals) that cannot be synthesized via vapor-liquid-solid syntheses.
\begin{acknowledgments}
This research was supported by grants from the National Science Foundation Division of Materials Research (\#1555007) and Air Force Office of Scientific Research Young Investigator Program (\#FA95501210400).
\end{acknowledgments}
\bibliographystyle{apsrev}
\bibliography{sinwdac}

\begin{thebibliography}{32}
\expandafter\ifx\csname natexlab\endcsname\relax\def\natexlab#1{#1}\fi
\expandafter\ifx\csname bibnamefont\endcsname\relax
  \def\bibnamefont#1{#1}\fi
\expandafter\ifx\csname bibfnamefont\endcsname\relax
  \def\bibfnamefont#1{#1}\fi
\expandafter\ifx\csname citenamefont\endcsname\relax
  \def\citenamefont#1{#1}\fi
\expandafter\ifx\csname url\endcsname\relax
  \def\url#1{\texttt{#1}}\fi
\expandafter\ifx\csname urlprefix\endcsname\relax\def\urlprefix{URL }\fi
\providecommand{\bibinfo}[2]{#2}
\providecommand{\eprint}[2][]{\url{#2}}

\bibitem[{\citenamefont{{Frederick K. Lutgens}}(2012)}]{lut12}
\bibinfo{author}{\bibnamefont{{Frederick K. Lutgens}}},
  \emph{\bibinfo{title}{Essentials of Geology}} (\bibinfo{publisher}{Prentice
  Hall}, \bibinfo{address}{Boston}, \bibinfo{year}{2012}),
  \bibinfo{edition}{11th} ed.

\bibitem[{\citenamefont{R\"odl et~al.}(2015)\citenamefont{R\"odl, Sander,
  Bechstedt, Vidal, Olsson, Laribi, and Guillemoles}}]{rod15}
\bibinfo{author}{\bibfnamefont{C.}~\bibnamefont{R\"odl}},
  \bibinfo{author}{\bibfnamefont{T.}~\bibnamefont{Sander}},
  \bibinfo{author}{\bibfnamefont{F.}~\bibnamefont{Bechstedt}},
  \bibinfo{author}{\bibfnamefont{J.}~\bibnamefont{Vidal}},
  \bibinfo{author}{\bibfnamefont{P.}~\bibnamefont{Olsson}},
  \bibinfo{author}{\bibfnamefont{S.}~\bibnamefont{Laribi}}, \bibnamefont{and}
  \bibinfo{author}{\bibfnamefont{J.-F.} \bibnamefont{Guillemoles}},
  \bibinfo{journal}{Physical Review B} \textbf{\bibinfo{volume}{92}},
  \bibinfo{pages}{045207} (\bibinfo{year}{2015}).

\bibitem[{\citenamefont{Wippermann et~al.}(2013)\citenamefont{Wippermann,
  V\"or\"os, Rocca, Gali, Zimanyi, and Galli}}]{wip13}
\bibinfo{author}{\bibfnamefont{S.}~\bibnamefont{Wippermann}},
  \bibinfo{author}{\bibfnamefont{M.}~\bibnamefont{V\"or\"os}},
  \bibinfo{author}{\bibfnamefont{D.}~\bibnamefont{Rocca}},
  \bibinfo{author}{\bibfnamefont{A.}~\bibnamefont{Gali}},
  \bibinfo{author}{\bibfnamefont{G.}~\bibnamefont{Zimanyi}}, \bibnamefont{and}
  \bibinfo{author}{\bibfnamefont{G.}~\bibnamefont{Galli}},
  \bibinfo{journal}{Physical Review Letters} \textbf{\bibinfo{volume}{110}},
  \bibinfo{pages}{046804} (\bibinfo{year}{2013}).

\bibitem[{\citenamefont{{Fontcuberta i Morral}
  et~al.}(2007)\citenamefont{{Fontcuberta i Morral}, Arbiol, Prades, Cirera,
  and Morante}}]{fon07}
\bibinfo{author}{\bibfnamefont{A.}~\bibnamefont{{Fontcuberta i Morral}}},
  \bibinfo{author}{\bibfnamefont{J.}~\bibnamefont{Arbiol}},
  \bibinfo{author}{\bibfnamefont{J.~D.} \bibnamefont{Prades}},
  \bibinfo{author}{\bibfnamefont{A.}~\bibnamefont{Cirera}}, \bibnamefont{and}
  \bibinfo{author}{\bibfnamefont{J.~R.} \bibnamefont{Morante}},
  \bibinfo{journal}{Advanced Materials} \textbf{\bibinfo{volume}{19}},
  \bibinfo{pages}{1347} (\bibinfo{year}{2007}).

\bibitem[{\citenamefont{Zhang et~al.}(1999)\citenamefont{Zhang, Iqbal,
  Vijayalakshmi, and Grebel}}]{zha99}
\bibinfo{author}{\bibfnamefont{Y.}~\bibnamefont{Zhang}},
  \bibinfo{author}{\bibfnamefont{Z.}~\bibnamefont{Iqbal}},
  \bibinfo{author}{\bibfnamefont{S.}~\bibnamefont{Vijayalakshmi}},
  \bibnamefont{and} \bibinfo{author}{\bibfnamefont{H.}~\bibnamefont{Grebel}},
  \bibinfo{journal}{Applied Physics Letters} \textbf{\bibinfo{volume}{75}},
  \bibinfo{pages}{2758} (\bibinfo{year}{1999}).

\bibitem[{\citenamefont{Kim et~al.}(2015)\citenamefont{Kim, Stefanoski,
  Kurakevych, and Strobel}}]{kim15}
\bibinfo{author}{\bibfnamefont{D.~Y.} \bibnamefont{Kim}},
  \bibinfo{author}{\bibfnamefont{S.}~\bibnamefont{Stefanoski}},
  \bibinfo{author}{\bibfnamefont{O.~O.} \bibnamefont{Kurakevych}},
  \bibnamefont{and} \bibinfo{author}{\bibfnamefont{T.~A.}
  \bibnamefont{Strobel}}, \bibinfo{journal}{Nature Materials}
  \textbf{\bibinfo{volume}{14}}, \bibinfo{pages}{169} (\bibinfo{year}{2015}).

\bibitem[{\citenamefont{Voronin et~al.}(2003)\citenamefont{Voronin, Pantea,
  Zerda, Wang, and Zhao}}]{vor03}
\bibinfo{author}{\bibfnamefont{G.~A.} \bibnamefont{Voronin}},
  \bibinfo{author}{\bibfnamefont{C.}~\bibnamefont{Pantea}},
  \bibinfo{author}{\bibfnamefont{T.~W.} \bibnamefont{Zerda}},
  \bibinfo{author}{\bibfnamefont{L.}~\bibnamefont{Wang}}, \bibnamefont{and}
  \bibinfo{author}{\bibfnamefont{Y.}~\bibnamefont{Zhao}},
  \bibinfo{journal}{Physical Review B} \textbf{\bibinfo{volume}{68}},
  \bibinfo{pages}{020102} (\bibinfo{year}{2003}).

\bibitem[{\citenamefont{Dyuzheva et~al.}(1978)\citenamefont{Dyuzheva,
  Kabalkina, and Novichkov}}]{dyu78}
\bibinfo{author}{\bibfnamefont{T.~I.} \bibnamefont{Dyuzheva}},
  \bibinfo{author}{\bibfnamefont{S.~S.} \bibnamefont{Kabalkina}},
  \bibnamefont{and} \bibinfo{author}{\bibfnamefont{V.~P.}
  \bibnamefont{Novichkov}}, \bibinfo{journal}{Zhurnal Eksperimentalnoi I
  Teoreticheskoi Fiziki} \textbf{\bibinfo{volume}{74}}, \bibinfo{pages}{1784}
  (\bibinfo{year}{1978}).

\bibitem[{\citenamefont{Hu and Spain}(1984)}]{hu84}
\bibinfo{author}{\bibfnamefont{J.~Z.} \bibnamefont{Hu}} \bibnamefont{and}
  \bibinfo{author}{\bibfnamefont{I.~L.} \bibnamefont{Spain}},
  \bibinfo{journal}{Solid State Communications} \textbf{\bibinfo{volume}{51}},
  \bibinfo{pages}{263} (\bibinfo{year}{1984}).

\bibitem[{\citenamefont{Hu et~al.}(1986)\citenamefont{Hu, Merkle, Menoni, and
  Spain}}]{hu86}
\bibinfo{author}{\bibfnamefont{J.~Z.} \bibnamefont{Hu}},
  \bibinfo{author}{\bibfnamefont{L.~D.} \bibnamefont{Merkle}},
  \bibinfo{author}{\bibfnamefont{C.~S.} \bibnamefont{Menoni}},
  \bibnamefont{and} \bibinfo{author}{\bibfnamefont{I.~L.} \bibnamefont{Spain}},
  \bibinfo{journal}{Physical Review B} \textbf{\bibinfo{volume}{34}},
  \bibinfo{pages}{4679} (\bibinfo{year}{1986}).

\bibitem[{\citenamefont{Tolbert and Alivisatos}(1995)}]{tol95}
\bibinfo{author}{\bibfnamefont{S.~H.} \bibnamefont{Tolbert}} \bibnamefont{and}
  \bibinfo{author}{\bibfnamefont{A.~P.} \bibnamefont{Alivisatos}},
  \bibinfo{journal}{Annual Review of Physical Chemistry}
  \textbf{\bibinfo{volume}{46}}, \bibinfo{pages}{595} (\bibinfo{year}{1995}).

\bibitem[{\citenamefont{Fabbri et~al.}(2013)\citenamefont{Fabbri, Rotunno,
  Lazzarini, Cavalcoli, Castaldini, Fukata, Sato, Salviati, and
  Cavallini}}]{fab13}
\bibinfo{author}{\bibfnamefont{F.}~\bibnamefont{Fabbri}},
  \bibinfo{author}{\bibfnamefont{E.}~\bibnamefont{Rotunno}},
  \bibinfo{author}{\bibfnamefont{L.}~\bibnamefont{Lazzarini}},
  \bibinfo{author}{\bibfnamefont{D.}~\bibnamefont{Cavalcoli}},
  \bibinfo{author}{\bibfnamefont{A.}~\bibnamefont{Castaldini}},
  \bibinfo{author}{\bibfnamefont{N.}~\bibnamefont{Fukata}},
  \bibinfo{author}{\bibfnamefont{K.}~\bibnamefont{Sato}},
  \bibinfo{author}{\bibfnamefont{G.}~\bibnamefont{Salviati}}, \bibnamefont{and}
  \bibinfo{author}{\bibfnamefont{A.}~\bibnamefont{Cavallini}},
  \bibinfo{journal}{Nano Letters} \textbf{\bibinfo{volume}{13}},
  \bibinfo{pages}{5900} (\bibinfo{year}{2013}).

\bibitem[{\citenamefont{Fabbri et~al.}(2014)\citenamefont{Fabbri, Rotunno,
  Lazzarini, Fukata, and Salviati}}]{fab14}
\bibinfo{author}{\bibfnamefont{F.}~\bibnamefont{Fabbri}},
  \bibinfo{author}{\bibfnamefont{E.}~\bibnamefont{Rotunno}},
  \bibinfo{author}{\bibfnamefont{L.}~\bibnamefont{Lazzarini}},
  \bibinfo{author}{\bibfnamefont{N.}~\bibnamefont{Fukata}}, \bibnamefont{and}
  \bibinfo{author}{\bibfnamefont{G.}~\bibnamefont{Salviati}},
  \bibinfo{journal}{Scientific Reports} \textbf{\bibinfo{volume}{4}}
  (\bibinfo{year}{2014}).

\bibitem[{\citenamefont{Besson et~al.}(1987)\citenamefont{Besson, Mokhtari,
  Gonzalez, and Weill}}]{bes87}
\bibinfo{author}{\bibfnamefont{J.~M.} \bibnamefont{Besson}},
  \bibinfo{author}{\bibfnamefont{E.~H.} \bibnamefont{Mokhtari}},
  \bibinfo{author}{\bibfnamefont{J.}~\bibnamefont{Gonzalez}}, \bibnamefont{and}
  \bibinfo{author}{\bibfnamefont{G.}~\bibnamefont{Weill}},
  \bibinfo{journal}{Physical Review Letters} \textbf{\bibinfo{volume}{59}},
  \bibinfo{pages}{473} (\bibinfo{year}{1987}).

\bibitem[{\citenamefont{Weill et~al.}(1989)\citenamefont{Weill, Mansot, Sagon,
  Carlone, and Besson}}]{wei89}
\bibinfo{author}{\bibfnamefont{G.}~\bibnamefont{Weill}},
  \bibinfo{author}{\bibfnamefont{J.~L.} \bibnamefont{Mansot}},
  \bibinfo{author}{\bibfnamefont{G.}~\bibnamefont{Sagon}},
  \bibinfo{author}{\bibfnamefont{C.}~\bibnamefont{Carlone}}, \bibnamefont{and}
  \bibinfo{author}{\bibfnamefont{J.~M.} \bibnamefont{Besson}},
  \bibinfo{journal}{Semiconductor Science and Technology}
  \textbf{\bibinfo{volume}{4}}, \bibinfo{pages}{280} (\bibinfo{year}{1989}).

\bibitem[{\citenamefont{Yablonovitch}(1993)}]{yab93}
\bibinfo{author}{\bibfnamefont{E.}~\bibnamefont{Yablonovitch}},
  \bibinfo{journal}{Journal of the Optical Society of America B}
  \textbf{\bibinfo{volume}{10}}, \bibinfo{pages}{283} (\bibinfo{year}{1993}).

\bibitem[{\citenamefont{Joannopoulos et~al.}(1997)\citenamefont{Joannopoulos,
  Villeneuve, and Fan}}]{joa97}
\bibinfo{author}{\bibfnamefont{J.~D.} \bibnamefont{Joannopoulos}},
  \bibinfo{author}{\bibfnamefont{P.~R.} \bibnamefont{Villeneuve}},
  \bibnamefont{and} \bibinfo{author}{\bibfnamefont{S.}~\bibnamefont{Fan}},
  \bibinfo{journal}{Nature} \textbf{\bibinfo{volume}{386}},
  \bibinfo{pages}{143} (\bibinfo{year}{1997}).

\bibitem[{\citenamefont{Birner et~al.}(2001)\citenamefont{Birner, Wehrspohn,
  G\"osele, and Busch}}]{bir01}
\bibinfo{author}{\bibfnamefont{A.}~\bibnamefont{Birner}},
  \bibinfo{author}{\bibfnamefont{R.~B.} \bibnamefont{Wehrspohn}},
  \bibinfo{author}{\bibfnamefont{U.~M.} \bibnamefont{G\"osele}},
  \bibnamefont{and} \bibinfo{author}{\bibfnamefont{K.}~\bibnamefont{Busch}},
  \bibinfo{journal}{Advanced Materials} \textbf{\bibinfo{volume}{13}},
  \bibinfo{pages}{377} (\bibinfo{year}{2001}).

\bibitem[{\citenamefont{Kennedy et~al.}(2003)\citenamefont{Kennedy, Brett,
  Miguez, Toader, and John}}]{ken03}
\bibinfo{author}{\bibfnamefont{S.~R.} \bibnamefont{Kennedy}},
  \bibinfo{author}{\bibfnamefont{M.~J.} \bibnamefont{Brett}},
  \bibinfo{author}{\bibfnamefont{H.}~\bibnamefont{Miguez}},
  \bibinfo{author}{\bibfnamefont{O.}~\bibnamefont{Toader}}, \bibnamefont{and}
  \bibinfo{author}{\bibfnamefont{S.}~\bibnamefont{John}},
  \bibinfo{journal}{Photonics and Nanostructures - Fundamentals and
  Applications} \textbf{\bibinfo{volume}{1}}, \bibinfo{pages}{37}
  (\bibinfo{year}{2003}).

\bibitem[{\citenamefont{Bermel et~al.}(2007)\citenamefont{Bermel, Luo, Zeng,
  Kimerling, and Joannopoulos}}]{ber07}
\bibinfo{author}{\bibfnamefont{P.}~\bibnamefont{Bermel}},
  \bibinfo{author}{\bibfnamefont{C.}~\bibnamefont{Luo}},
  \bibinfo{author}{\bibfnamefont{L.}~\bibnamefont{Zeng}},
  \bibinfo{author}{\bibfnamefont{L.~C.} \bibnamefont{Kimerling}},
  \bibnamefont{and} \bibinfo{author}{\bibfnamefont{J.~D.}
  \bibnamefont{Joannopoulos}}, \bibinfo{journal}{Optics Express}
  \textbf{\bibinfo{volume}{15}}, \bibinfo{pages}{16986} (\bibinfo{year}{2007}).

\bibitem[{\citenamefont{Roder et~al.}(2012)\citenamefont{Roder, Pauzauskie, and
  Davis}}]{rod12}
\bibinfo{author}{\bibfnamefont{P.~B.} \bibnamefont{Roder}},
  \bibinfo{author}{\bibfnamefont{P.~J.} \bibnamefont{Pauzauskie}},
  \bibnamefont{and} \bibinfo{author}{\bibfnamefont{E.~J.} \bibnamefont{Davis}},
  \bibinfo{journal}{Langmuir} \textbf{\bibinfo{volume}{28}},
  \bibinfo{pages}{16177} (\bibinfo{year}{2012}).

\bibitem[{\citenamefont{Garnett and Yang}(2010)}]{gar10}
\bibinfo{author}{\bibfnamefont{E.}~\bibnamefont{Garnett}} \bibnamefont{and}
  \bibinfo{author}{\bibfnamefont{P.}~\bibnamefont{Yang}},
  \bibinfo{journal}{Nano Letters} \textbf{\bibinfo{volume}{10}},
  \bibinfo{pages}{1082} (\bibinfo{year}{2010}).

\bibitem[{\citenamefont{Huang et~al.}(2011)\citenamefont{Huang, Geyer, Werner,
  de~Boor, and G\"osele}}]{hua11}
\bibinfo{author}{\bibfnamefont{Z.}~\bibnamefont{Huang}},
  \bibinfo{author}{\bibfnamefont{N.}~\bibnamefont{Geyer}},
  \bibinfo{author}{\bibfnamefont{P.}~\bibnamefont{Werner}},
  \bibinfo{author}{\bibfnamefont{J.}~\bibnamefont{de~Boor}}, \bibnamefont{and}
  \bibinfo{author}{\bibfnamefont{U.}~\bibnamefont{G\"osele}},
  \bibinfo{journal}{Advanced Materials} \textbf{\bibinfo{volume}{23}},
  \bibinfo{pages}{285} (\bibinfo{year}{2011}).

\bibitem[{\citenamefont{Smith et~al.}(2015)\citenamefont{Smith, Roder, Hanson,
  Manandhar, Devaraj, Perea, Kim, Kilcoyne, and Pauzauskie}}]{smi15}
\bibinfo{author}{\bibfnamefont{B.~E.} \bibnamefont{Smith}},
  \bibinfo{author}{\bibfnamefont{P.~B.} \bibnamefont{Roder}},
  \bibinfo{author}{\bibfnamefont{J.~L.} \bibnamefont{Hanson}},
  \bibinfo{author}{\bibfnamefont{S.}~\bibnamefont{Manandhar}},
  \bibinfo{author}{\bibfnamefont{A.}~\bibnamefont{Devaraj}},
  \bibinfo{author}{\bibfnamefont{D.~E.} \bibnamefont{Perea}},
  \bibinfo{author}{\bibfnamefont{W.-J.} \bibnamefont{Kim}},
  \bibinfo{author}{\bibfnamefont{A.~L.~D.} \bibnamefont{Kilcoyne}},
  \bibnamefont{and} \bibinfo{author}{\bibfnamefont{P.~J.}
  \bibnamefont{Pauzauskie}}, \bibinfo{journal}{ACS Photonics}
  \textbf{\bibinfo{volume}{2}}, \bibinfo{pages}{559} (\bibinfo{year}{2015}).

\bibitem[{\citenamefont{Mao et~al.}(1986)\citenamefont{Mao, Xu, and
  Bell}}]{mao86}
\bibinfo{author}{\bibfnamefont{H.~K.} \bibnamefont{Mao}},
  \bibinfo{author}{\bibfnamefont{J.}~\bibnamefont{Xu}}, \bibnamefont{and}
  \bibinfo{author}{\bibfnamefont{P.~M.} \bibnamefont{Bell}},
  \bibinfo{journal}{Journal of Geophysical Research: Solid Earth}
  \textbf{\bibinfo{volume}{91}}, \bibinfo{pages}{4673} (\bibinfo{year}{1986}).

\bibitem[{\citenamefont{Kittel}(2005)}]{kit05}
\bibinfo{author}{\bibfnamefont{C.}~\bibnamefont{Kittel}},
  \emph{\bibinfo{title}{Introduction to {Solid} {State} {Physics}}}
  (\bibinfo{publisher}{John Wiley and Sons}, \bibinfo{year}{2005}),
  \bibinfo{edition}{8th} ed.

\bibitem[{\citenamefont{Khachadorian et~al.}(2013)\citenamefont{Khachadorian,
  Papagelis, Ogata, Hofmann, Phillips, and Thomsen}}]{kha13}
\bibinfo{author}{\bibfnamefont{S.}~\bibnamefont{Khachadorian}},
  \bibinfo{author}{\bibfnamefont{K.}~\bibnamefont{Papagelis}},
  \bibinfo{author}{\bibfnamefont{K.}~\bibnamefont{Ogata}},
  \bibinfo{author}{\bibfnamefont{S.}~\bibnamefont{Hofmann}},
  \bibinfo{author}{\bibfnamefont{M.~R.} \bibnamefont{Phillips}},
  \bibnamefont{and} \bibinfo{author}{\bibfnamefont{C.}~\bibnamefont{Thomsen}},
  \bibinfo{journal}{The Journal of Physical Chemistry C}
  \textbf{\bibinfo{volume}{117}}, \bibinfo{pages}{4219} (\bibinfo{year}{2013}).

\bibitem[{\citenamefont{Weinstein and Piermarini}(1975)}]{wei75}
\bibinfo{author}{\bibfnamefont{B.~A.} \bibnamefont{Weinstein}}
  \bibnamefont{and} \bibinfo{author}{\bibfnamefont{G.~J.}
  \bibnamefont{Piermarini}}, \bibinfo{journal}{Physical Review B}
  \textbf{\bibinfo{volume}{12}}, \bibinfo{pages}{1172} (\bibinfo{year}{1975}).

\bibitem[{\citenamefont{Giannozzi et~al.}(2009)\citenamefont{Giannozzi, Baroni,
  Bonini, Calandra, Car, Cavazzoni, Ceresoli, Chiarotti, Cococcioni, Dabo
  et~al.}}]{gia09}
\bibinfo{author}{\bibfnamefont{P.}~\bibnamefont{Giannozzi}},
  \bibinfo{author}{\bibfnamefont{S.}~\bibnamefont{Baroni}},
  \bibinfo{author}{\bibfnamefont{N.}~\bibnamefont{Bonini}},
  \bibinfo{author}{\bibfnamefont{M.}~\bibnamefont{Calandra}},
  \bibinfo{author}{\bibfnamefont{R.}~\bibnamefont{Car}},
  \bibinfo{author}{\bibfnamefont{C.}~\bibnamefont{Cavazzoni}},
  \bibinfo{author}{\bibfnamefont{D.}~\bibnamefont{Ceresoli}},
  \bibinfo{author}{\bibfnamefont{G.~L.} \bibnamefont{Chiarotti}},
  \bibinfo{author}{\bibfnamefont{M.}~\bibnamefont{Cococcioni}},
  \bibinfo{author}{\bibfnamefont{I.}~\bibnamefont{Dabo}}, \bibnamefont{et~al.},
  \bibinfo{journal}{Journal of Physics: Condensed Matter}
  \textbf{\bibinfo{volume}{21}}, \bibinfo{pages}{395502}
  (\bibinfo{year}{2009}).

\bibitem[{\citenamefont{Khachadorian et~al.}(2011)\citenamefont{Khachadorian,
  Papagelis, Scheel, Colli, Ferrari, and Thomsen}}]{kha11}
\bibinfo{author}{\bibfnamefont{S.}~\bibnamefont{Khachadorian}},
  \bibinfo{author}{\bibfnamefont{K.}~\bibnamefont{Papagelis}},
  \bibinfo{author}{\bibfnamefont{H.}~\bibnamefont{Scheel}},
  \bibinfo{author}{\bibfnamefont{A.}~\bibnamefont{Colli}},
  \bibinfo{author}{\bibfnamefont{A.~C.} \bibnamefont{Ferrari}},
  \bibnamefont{and} \bibinfo{author}{\bibfnamefont{C.}~\bibnamefont{Thomsen}},
  \bibinfo{journal}{Nanotechnology} \textbf{\bibinfo{volume}{22}},
  \bibinfo{pages}{195707} (\bibinfo{year}{2011}).

\bibitem[{\citenamefont{Hsieh}(2015)}]{hsi15}
\bibinfo{author}{\bibfnamefont{W.-P.} \bibnamefont{Hsieh}},
  \bibinfo{journal}{Journal of Applied Physics} \textbf{\bibinfo{volume}{117}},
  \bibinfo{pages}{235901} (\bibinfo{year}{2015}).

\bibitem[{\citenamefont{Goncharov et~al.}(2012)\citenamefont{Goncharov, Wong,
  Dalton, Ojwang, Struzhkin, Kon™pkov‡, and Lazor}}]{gon12}
\bibinfo{author}{\bibfnamefont{A.~F.} \bibnamefont{Goncharov}},
  \bibinfo{author}{\bibfnamefont{M.}~\bibnamefont{Wong}},
  \bibinfo{author}{\bibfnamefont{D.~A.} \bibnamefont{Dalton}},
  \bibinfo{author}{\bibfnamefont{J.~G.~O.} \bibnamefont{Ojwang}},
  \bibinfo{author}{\bibfnamefont{V.~V.} \bibnamefont{Struzhkin}},
  \bibinfo{author}{\bibfnamefont{Z.}~\bibnamefont{Kon™pkov‡}},
  \bibnamefont{and} \bibinfo{author}{\bibfnamefont{P.}~\bibnamefont{Lazor}},
  \bibinfo{journal}{Journal of Applied Physics} \textbf{\bibinfo{volume}{111}},
  \bibinfo{pages}{112609} (\bibinfo{year}{2012}).

\end{thebibliography}

\end{document}